\documentclass[iop]{emulateapj}

\def\simgreat{\mathbin{\lower 3pt\hbox
     {$\rlap{\raise 5pt\hbox{$\char'076$}}\mathchar"7218$}}}
\def\simless{\mathbin{\lower 3pt\hbox
     {$\rlap{\raise 5pt\hbox{$\char'074$}}\mathchar"7218$}}}

\usepackage{amsmath}
\usepackage{natbib}
\bibpunct{(}{)}{;}{a}{}{,}
\usepackage{graphicx}
\usepackage{appendix}
\begin{document}
\title{GPU--based Monte Carlo dust radiative transfer scheme applied
  to AGN} \author{Frank Heymann}
\affil{Department of Physics and Astronomy, University of Kentucky,
  Lexington, KY 40506-0055, USA }
\affil{European Southern Observatory (ESO), Karl-Schwarzschild-Str. 2,
  85748 Garching, Germany}
\affil{Astronomisches Institut, Ruhr-University, Bochum, Germany}
\email{fheymann@pa.uky.edu}
\and
\author{Ralf Siebenmorgen}
\affil{European Southern Observatory (ESO), Karl-Schwarzschild-Str. 2,
  85748 Garching, Germany\\}
\begin{abstract} A three dimensional parallel Monte Carlo (MC) dust
  radiative transfer code is presented. To overcome the huge computing
  time requirements of MC treatments, the computational power of
  vectorized hardware is used, utilizing either multi--core computer
  power or graphics processing units. The approach is a
  self-consistent way to solve the radiative transfer equation in
  arbitrary dust configurations.  The code calculates the equilibrium
  temperatures of two populations of large grains and stochastic
  heated polycyclic aromatic hydrocarbons (PAH).  Anisotropic
  scattering is treated applying the Heney--Greenstein phase
  function. The spectral energy distribution (SED) of the object is
  derived at low spatial resolution by a photon counting procedure and
  at high spatial resolution by a vectorized ray--tracer. The latter
  allows computation of high signal--to--noise images of the objects
  at any frequencies and arbitrary viewing angles.  We test the
  robustness of our approach against other radiative transfer
  codes. The SED and dust temperatures of one and two dimensional
  benchmarks are reproduced at high precision. The parallelization
  capability of various MC algorithms is analyzed and included in our
  treatment.  We utilize the Lucy--algorithm for the optical thin case
  where the Poisson noise is high, the iteration free Bjorkman \& Wood
  method to reduce the calculation time, and the Fleck \& Canfield
  diffusion approximation for extreme optical thick cells.  The code
  is applied to model the appearance of active galactic nuclei (AGN)
  at optical and infrared wavelengths. The AGN torus is clumpy and
  includes fluffy composite grains of various sizes made--up of
  silicates and carbon. The dependence of the SED on the number of
  clumps in the torus and the viewing angle is studied. The appearance
  of the 10$\mu$m silicate features in absorption or emission is
  discussed. The SED of the radio loud quasar 3C~249.1 is fit by the
  AGN model and a cirrus component to account for the far infrared
  emission.
\end{abstract}
\section{Introduction}
\label{sec:introduction}
Dust obscured objects cannot be studied directly in the UV/optical,
since the dust shields most of the visible light. To derive the
UV/optical component or to constrain the morphological structure from
available near/far infrared data, a detailed model of the interaction
of photons with the dust is required.  This necessitates a solution to
the radiative transfer (RT) equation.  Analytically, this can only be
done in some simple configurations, for example by assuming spherical
or disk symmetry in which either scattering or absorption is neglected
and the wavelength dependency of the dust cross section is strongly
simplified, e.g. by a gray body approximation. Nature, however, is
usually not well approximated by such assumptions and numerical
modeling is the only way to solve this problem.

Dust is detected in the majority of active galactic nuclei
\citep{haa08}.  According to the unified scheme \citep{ant85}, AGN are
surrounded by a dust obscuring torus.  This torus, as argued by
\cite{kro88}, needs a clumpy structure to allow the survival of dust
grains in regions where the gas \hbox{temperatures ($\sim10^6$\,K)}
are extreme compared to the dust sublimation temperature.  Indeed,
\cite{tri07} show with VLTI observations of the Circinus active
galactic nuclei (AGN) strong evidence for a clumpy or filamentary
structure of the nucleus.

In this paper a numerical method is presented, which solves the
radiative transfer equation in a three dimensional geometry, based on
the MC technique (\citealt{wit77}; Sect.\ref{sec:MCRT}).  To reduce
the required computational effort, different optimization strategies
are developed
(\citealt{luc99,bw01,gor01,mis01,wol03,bae08,bia08,bae11};
Sect.~\ref{sec:optimizations}). Our numerical solution of the
radiative transfer equation takes advantage of some of these
optimization algorithms and is specifically developed to be vectorized
and to run on graphics processing units (GPU).  They are introduced
into the original code developed by \cite{kru08}. The MC routine
handles arbitrary dust distributions in a three dimensional Cartesian
model space at various optical depths. The self-consistent solution
provides the dust temperatures; spectral energy distributions (SED)
and images are computed using a ray--tracer
(Sect.~\ref{sec:raytracing}).  The code is tested against existing
benchmark results (Sect.~\ref{sec:benchmark}).  The influence of
clumpiness of an AGN dust torus on the SED and the 10$\mu$m silicate
feature is discussed and a model of the radio loud quasar 3C249.1 is
presented (Sect.~\ref{sec:clumpy_torus_model}).

\section{Monte Carlo Radiative Transfer}
\label{sec:MCRT}
\begin{figure}[t]
  \includegraphics[width=0.5\textwidth]{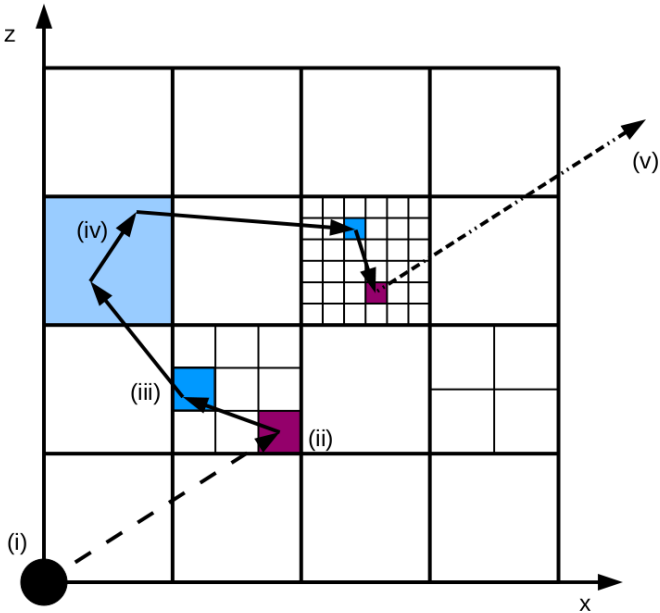}
  \caption {2D illustration of the three dimensional grid and the
    trajectories of the photon packets. The grid is divided into
    cubes, which can be further divided into sub-cubes. The source
    emits photons at (i), which interact or not with the dust. By
    interaction photon packets are either absorbed (ii) or scattered
    (iii). Multiple interactions may occur in one cell (iv). After
    absorption the packet will be re-emitted by the dust at a
    different frequency.  Frequency changes are illustrated by
    different line styles. Finally, (v) the packet escapes the model
    space.}
  \label{fig:grid}
\end{figure}

In our MC procedure the bolometric luminosity
$L=\nolinebreak\int_0^\infty L_\nu d\nu$ of the central engine is
divided into $N=m ~ n_{\rm{zyk}}$ monochromatic photon packets of
equal energy $\xi=L/N$, where $m$ is the total number of frequency
bins of the heating source and $n_{\rm{zyk}}$ counts the number of
photon packets per bin. The width of the frequency bin with index j
between the frequencies $\nu_j$ and $\nu_{j+1}$, $\Delta \nu =
|\nu_{j+1} - \nu_j|$, is derived from the spectral shape of the
radiation source and the energy of a photon packet
\begin{equation}
  \xi = \int_{\displaystyle \nu_j}^{\displaystyle \nu_{j+1}}L_\nu d\nu ~.
  \label{eqn::freq_grid}
\end{equation}

The primary heating source emits photons for example in the optical
and UV while the dust emission occurs at IR and sub-millimeter
wavelength. Two frequency grids are considered: The first one is set
up by Eq. \eqref{eqn::freq_grid} and accounts for the emission of the
primary heating source. The second frequency grid is an extended
version of the first one and includes additional bins in the IR and
sub-millimeter to scope with the dust emission. The interaction
probability between photons and dust particles in the radiative
transfer problem is computed by the absorption $\kappa^{abs}_\nu$ and
scattering $\kappa^{sca}_\nu$ cross-sections. The code includes
different grain materials and size distributions.

The model space is set up as cubes in an orthogonal Cartesian grid.
Each cube is divided into an arbitrary number of sub-cubes of volume
$V_i$ and constant density $\rho_i$ (Fig. \ref{fig:grid}), where the
index refers to sub-cube $i$.  The subdivision of cubes allows a finer
sampling whenever required: For example close to the dust evaporation
zone, in regions of high optical depth or where the spatial gradient
of the radiation field is large. One possible trajectory of a photon
is illustrated in Fig. \ref{fig:grid}.  Photons are emitted by the
source at frequency $\nu_{j_{\star}}$, where $j_{\star}$ is the index
of the frequency grid of the source. The flight path of the photon
through out the model space is computed.

The direction of the photon is chosen from a uniform distribution of
random numbers $\zeta_1,\zeta_2$ from the half open interval $[0,1)$,
so that $\phi = 2\pi\zeta_1$ and $\cos{\theta} = 1-2\zeta_2$. The
distance from the entry point of a photon into a cell $i$ to its exit
point along the travel direction is $\ell_i$.  In the MC method the
interaction of photons with the dust in a sub-cube can be determined
with an uniform distributed random number $\zeta$ \citep{wit77,luc99}
and the optical depth

\begin{equation}
  \tau_i(\nu) = (\kappa^{abs}_\nu+\kappa^{sca}_\nu)\ \rho_i\ \ell_i ~.
\end{equation}

Photons leave the cell if $\tau_i(\nu)\le-\log{\zeta}$ and otherwise
they interact with the dust.  If a packet passes through a cell
without interacting, it enters a neighboring cell according to its
direction of travel. At the border of the model space it eventually
escapes. When a photon packet enters a cell a new random number is
chosen to determine the interaction probability of photons with the
dust \footnote{It can be confirmed that the interaction probability
  calculated with a new random number is in agreement with the
  treatment of a travel distance which is determined by a single
  random number (\cite{luc99}).}. The photon packet interacts with the
dust after it has traveled a distance

\begin{equation}
  \ell_i' = \frac{-\log{\zeta}}{\rho_i(\kappa^{abs}_\nu+\kappa^{sca}_\nu)} ~.
\end{equation}

The travel distance $\ell_i'$ defines the point in the cell where
interaction takes place and photons are either scattered
(Fig. \ref{fig:grid}.ii) or absorbed (Fig. \ref{fig:grid}.iii).
Depending on its optical depth multiple scattering or absorption
events may occur in a cell (Fig.~\ref{fig:grid}.iv). In case of an
interaction event the probability of scattering is given by the albedo
$A_\nu=\kappa^{sca}_\nu\ /\ (\kappa^{sca}_\nu+\kappa^{abs}_\nu)$ and
therefore the chance of absorption is $1-A_\nu$.

When a photon is scattered on a dust grain the packet keeps its
frequency, but changes its travel direction according to the phase
function of the particle. We use the \cite{HG41} phase function to
approximate the anisotropic scattering (see
Sect. \ref{sec:raytracing}).  In the isotropic case the asymmetry
factor $g_\nu$ equals 0. The index of the cell, the frequency of the
scattered photon and the incoming direction is stored. This allows
computing of the source function of the cell $i$, which will be used
in a ray--tracer developed to compute the observed image at any
frequency (Sect.~\ref{sec:raytracing}).

When a photon is absorbed a new one with different frequency and
direction is emitted from that position. The frequency of the emitted
photon packet is given by the temperature of the absorbing
material. The emitted photon has the same energy as the absorbed
one. All photon packets contain the same amount of energy $\xi$,
therefore the total number of absorbed photon packets per particle and
cell is used to calculate the temperature of each material. After $k$
absorptions of photon packets by dust the cell $i$ has a temperature
$T_{i,k}$ calculated from:

\begin{equation}
  \int{\kappa^{abs}_\nu B_\nu(T_{i,k})\ d\nu}\ =\ \frac{k\xi}{4\pi \rho_i V_i} ~,
  \label{eqn:calc_temp}
\end{equation}
where $B_\nu(T_{i,k})$ is the Planck function.

Dust emits a photon packet at the shortest frequency $\nu'$ calculated
with:

\begin{equation}
  \int_0^{\nu'}{\kappa^{abs}_\nu\ B_\nu(T_{i,k})\ d\nu}\ \ge\ \zeta \int_0^{\infty}{\kappa^{abs}_\nu\ B_\nu(T_{i,k})\ d\nu} ~.
  \label{eqn:freq_calc1}
\end{equation}

In the MC method one follows all $N$ photon packets through the dust
cloud until they reach the outer boundary of the model and escape.
For each cell the absorption events are counted. When all packets
escaped the model space, the number of absorptions in a cell is used
(Eq.~\eqref{eqn:calc_temp}) to calculate its dust temperature.  An
iterative MC method is started with an initial guess of the dust
temperatures.  This allows computing of the frequencies of packets
emitted by the dust (Eq.~\eqref{eqn:freq_calc1}).  After all packets
escape the model space another $N$ photons are emitted by the source
and their flight paths through the model are computed. However, this
time using the dust temperatures of the previous run. This procedure
continues until the dust temperatures converges, which usually takes
about 3--5 iterations.

\subsection{Non equilibrium radiative transfer (PAH)}
The enthalpy state of very small grains such as polycyclic aromatic
hydrocarbons \citep{tie08} fluctuates strongly when they are hit by
photons. In order to solve the radiative transfer problem including
PAH with MC we treat their emission as a stochastic process. A
temperature distribution function $P(T)$ of the PAH is computed using
the iterative scheme by Siebenmorgen et al. (1992).  PAH are
implemented into the MC code in two steps: first, we compute the PAH
absorption of photon packages and store position and frequency of
these events. Then, after calculating the equilibrium temperature of
the large grains (Sect. \ref{sec:MCRT}), the PAH emission is
computed. Each cell with PAH absorption becomes a source and emits the
same number of photon packages as previously absorbed by the
PAH. These photons are traced through the model as described in
Sect. \ref{sec:MCRT}. The frequency of the emitted package is
calculated according to:
\begin{equation}
  \int_0^{\nu'}  \epsilon_n(\nu) d\nu \ge \xi \int \epsilon_n(\nu) d\nu \
\end{equation}
with random number $\xi$ and PAH emissivity
\begin{equation}
  \epsilon_n(\nu) = K^{\rm {PAH}}_\nu \int B_\nu(T) P_n(T) dT \ ,
\end{equation}
where $K^{\rm {PAH}}_\nu$ is the PAH cross-section per unit mass of
dust and $P_n(T)$ is the temperature distribution function as
described in \cite{sie10}.

Similar to the evaporation of large grains it is important to compute
the location where PAH are able to survive the radiation
environment. We treat the photo--dissociation of PAH as discussed in
\cite{sie10}.  They find that PAH dissociate when, within a cooling
time, the total absorbed energy $E_{\rm{PAH}}$ is larger than a
critical energy $E_{\rm{crit}}:$
\begin{equation}
  E_{\rm{PAH}} \ge E_{\rm {crit}} = \frac{N_{\rm C}}{2} \/ \ {\rm {(eV)}},
  \label{pahdestruct}
\end{equation}
where $N_{\rm{C}}$ is the number of carbon atoms per PAH molecule.  In
cells where the molecules dissociate the absorbed photon packages are
treated as if they are emitted from the star.  Further details on our
and other numerical implementations of PAH in the MC code, a
verification against benchmark models and an application to the PAH
emission from proto--planetary disks is given in an accompanying paper
\citep{sie12}.

\section{Optimized Monte Carlo radiative transfer}
\label{sec:optimizations}

In this section we describe various optimization techniques which are
introduced into our MC scheme to increase the accuracy and decrease
computing times.

\subsection{Small optical depth}
\label{sub:small-optical-depth}
The MC approach solves the radiative transfer problem in a
probabilistic way. Dust temperatures become uncertain and have large
errors whenever the interaction probability is low. This is the case
in cells of low optical depth where many photons fly--by without being
absorbed or scattered by the dust. The statistical noise in such cells
can be reduced significantly applying a procedure developed by
\cite{luc99}. Contrary to the method described in
Sect.~\ref{sec:MCRT}, in which only absorption events contribute to
the temperature calculation of the cell, the Lucy method considers the
absorbed energy of every photon packet traveling through the
cell. This technique increases the temperature accuracy of the MC
treatment in low optical depth regions. It has no advantage when the
optical depth $\tau_i \ge 1$ as interactions between photons and dust
are likely.

We follow Sect.3.4 of \cite{luc99} and implement a similar
optimization technique. It is applied whenever the maximal optical
depth of a cell $i$ is smaller than $10^{-2}$. The performance of the
MC scheme depends critically on this value.  It is chosen so that the
performance of the code is high while still accounting for reasonable
accuracy.  For smaller values the optimization is only applied for
extreme optical thin cells whereas for higher values the expensive
accuracy enhancement is applied to cells which can be accurately
computed without the Lucy scheme.  With our choice the energy absorbed
by the dust $E^{\rm {abs}}_{i}$ is a small fraction of the energy of a
photon packet. The energy is computed by

\begin{equation}
  E^{\rm {abs}}_{\rm i} = (1-e^{-\tau_{\rm i}}) \, \xi
\end{equation}

where $\tau_{\rm i}$ is the optical depth from the entry to the exit
point of the photon of cell $i$. Thereby the frequency of the photon
packet remains unchanged.

\subsection{High optical depth}
\label{sub:high-optical-depth}
The MC solution of the radiative transfer equation becomes slow for
very optical thick regions, where the number of photon interactions
with the dust increases exponentially. To avoid photons becoming
trapped in cells with very high optical depth we apply a modified
random walk procedure. It is based on a diffusion approximation by
\cite{fle84}. The method is tested by \cite{min09}
and numerical enhancements are given by \cite{rob10}. \\

\subsection{Iteration--free Monte Carlo}
\label{sub:iteration-free-monte}

An iteration free MC scheme is developed by \cite{bw01} (B\&W); see
\cite{bae05b,kru08} for helpful comments. In the B\&W method the dust
temperature in a cell is adjusted with respect to the previous
absorption event. Contrary to Eq.~\eqref{eqn:freq_calc1}, dust
re-emission occurs at the shortest frequency $\nu'$ for which

\begin{equation}
  \int_0^{\nu'}{\kappa^{abs}_\nu \frac{dB_\nu(T_{i,k})}{dT} d\nu}\ \ge\ \zeta \int_0^{\infty}{\kappa^{abs}_\nu \frac{dB_\nu(T_{i,k})}{dT} \ d\nu} 
  \label{eqn:BW}
\end{equation}

is valid. Both methods described in Eq.~\eqref{eqn:freq_calc1} and
Eq.~\eqref{eqn:BW} are implemented in the MC code and can be used upon
preference.

\subsection{Parallelization on graphics processing units}
\label{sub:parall-graph-proc}

The MC method of solving the radiative transfer problem is
particularly suited for parallelization because all photon packets are
independent of each other \citep{jon06,hey10,sie11}.  Vectorization of
the code allows a number of photon packets to be emitted
simultaneously. The number of packets launched at a time depends on
the number of threads\footnote{A thread is the smallest unit of
  processing that can be scheduled in a parallel environment.}
available on the computer. The trajectories of the photons can then be
calculated in parallel.

In parallel environments an additional requirement is placed on the
random number generator; each thread must have an independent sequence
of random numbers. We solve this problem by applying a parallel
version of the Mersenne Twister algorithm as given by \cite{mat00}.

Parallelization is most efficient when all processing units finish
their task at about the same time. Then vectorization speeds up the
code roughly equal to the number of threads available.  Unfortunately
this is not always the case.  The number of interactions of photons
increases exponentially with the optical depth $\tau_{\rm{V}}$ and
photon packets may get trapped in cells with say $\tau_{\rm{V}} \geq
1000$. The workload of these cells is much higher than for cells at
much lower optical depth. This results in a rather unbalanced workload
over all threads so that the advantage of vectorization may
disappear. In our application idle threads are avoided as much as
possible. Every time a thread finishes the trajectory of the photon
packet within the model space a counter is increased.  If the counter
reaches $80\%$ of the total number of parallel working threads and if
the number of interactions of the photon packet within a cell is
larger than 100 the thread pauses.  In this case the position and
frequency of the photon packet is stored. Close to the end of the
simulation when all photons with average processing speed have escaped
the model space, the stored photons are resumed.  This procedure
allows a balanced workload among all threads. In addition the modified
random walk procedure of \cite{rob10} is implemented
(Sect.~\ref{sub:high-optical-depth}.)

This code utilizes two different parallelization methods. It can be
used on shared memory machines using the openMP library to run as many
parallel rays as there are processor cores available and on vectorized
hardware (GPU) highly optimized for parallelization. The complete MC
radiative transfer solution is ported to GPU using the Compute Unified
Device Architecture (CUDA) developed by NVIDIA \citep{cuda}. This
speeds up the entire MC solution in comparison to the method by
\cite{jon10}, which provides only a GPU acceleration when computing
the temperature of a grain.

\subsection{Numerical recipes}

For vectorized MC radiative transfer codes it is recommended to follow
some general recipes.  In vectorized systems the increased number of
processing units decreases the computing time for numerical operations
whereas the time needed for read--and--write operations remains
unaltered.  This implies a paradigm shift in the programming
strategy. In single processing codes the performance is often improved
by reading pre--calculated values from memory.  Instead in a
vectorized code at some point it is faster to calculate such values
case by case.  For example in our code, and as long as the dust
frequency grid has less than $300$ bins, it is faster to solve
Eq. (\ref{eqn:freq_calc1}) on the fly than reading pre--computed
values from memory.

The performance of parallel codes is increased considerably when
designing a particular array structure for the particular hardware in
use. A read operation from memory is most efficient when it accesses
the entire array. For the GPUs used in this paper the memory bandwidth
is 512\,bit and we use a 32\,bit machine.  Therefore in our code the
most efficient way to read from memory is by a modulus of $16 \times
32$\,bit values. The design of a particular array structure is import
for problems where the memory access is not predictable, which is
unfortunately the case in Monte Carlo schemes.

In order to improve the performance it is very efficient to code with
as less interference between workers and threads as possible. The code
should avoid situations where thread A waits until thread B has
finished and then thread B waits until thread A has
finished. Unfortunately in MC simulations the dependency between
workers and threads is often less obvious than in the previous
example.

Another challenge with memory interactions is to use local copies of
arrays for each thread as often as possible. This decreases the
interaction between different threads and usually leads to large
performance improvements, of course at the expense of some additional
memory needs. When dealing with memory interactions between threads it
is helpful to study the special hardware capabilities. In our case we
gain a factor 2 in the computing time by using atomic integer
operations rather than floating point operations.

It is of advantage to optimize all read--and--write operations to the
fastest available source. This can be done by caching small amounts of
data from the disk to the global memory, than to the GPU memory and
finally to the GPU shared memory. For example in our code we store the
wavelength dependent cross--sections in the GPU shared memory. This is
about 10 times faster than reading them from the global memory.

For MC simulations the random number generator (RNG) is important and
special care must be taken on parallel systems.  A detailed study of
this topic is beyond the scope of this paper.  We tested different RNG
available and find that the performance and accuracy as well as thread
dependence differs significantly between various RNG
implementations. It is therefor important to choose and test a RNG
which fits the particular problem best.

\subsection{Parallelization and optimization algorithms}

\begin{table}[t]
  \caption{Run time requirements of different MC methods.}
  \begin{center}
    \begin{tabular}{|l |r |r |}
      \hline\hline
      Method & Threads & Time \\
      &  & (min.) \\
      \hline
      Lucy  & $1$ &  180 \\
      B\&W & $1$    &   60 \\
      Lucy  & $8$ & 45  \\
      B\&W & $8$    &  20 \\
      GPU & $256$    & 3 \\
      GPU+Lucy & $256$   & 4  \\
      GPU+B\&W& $256$ &   2 \\
      GPU+B\&W+Lucy& $256$ &   2 \\
      \hline
    \end{tabular}
    \label{tab:comp_MC_opt}
  \end{center}
\end{table}

The run time requirements of the different MC methods are compared. We
consider a star at temperature of 2500\,K with solar luminosity which
heats a spherical and constant density dust envelope with an inner
radius of 0.7\,AU and an outer radius of 700\,AU.  The optical depth
measured from the star to the outer boundary is
$\tau_{\rm{V}}=10$. Parameters of the dust are specified as in
Sect.~\ref{sub:bench1d}.  In the models $10^7$ photon packets are
emitted from the star and $10^6$ grid cells are used.

We apply the Lucy (Sect.3.1), B\&W (Sect.3.3.) method in a scalar
(single thread), a CPU version with 8 threads, a GPU version with 256
threads and in addition the iterative MC scheme of
Sect.~\ref{sec:MCRT} on a GPU machine.  Initializing times of the
various MC methods are identical.

As described in Sect.3.1 for the same number of emitted photon packets
the Lucy method provides a better temperature estimate than the other
methods. In the Lucy method the fraction of a photon packet is
considered.  This can only be realized by considering floating point
operations which are more computer expensive than integer
operations. In our test case the scalar version of the Lucy method is
run with three iterations and requires a total run time of 3 hours
(Tab.~\ref{tab:comp_MC_opt}).

For single thread applications the iteration--free MC scheme by B\&W
(Sect.~\ref{sub:iteration-free-monte}) has the advantages that it
speeds up the process by a factor which equals the number of
iterations required for convergence in the other MC treatments.
However, in the B\&W method memory interaction between different cells
are unavoidable and they slow down the parallelization capabilities of
this method. The memory interaction between cells rises with the
number of threads and therefore the amount of necessary {\it {atomic
    memory operations}}\footnote{Atomic operations are operations
  which are performed without interference from any other
  threads. Atomic operations are often used to prevent conditions
  which are common problems in multi--thread applications.} increases
with parallelization. To minimize the number of atomic memory
operations we solve the problem in an iterative MC scheme using
Eq.~\eqref{eqn:calc_temp}. This allows parallelization with a huge
number of threads. At low budget this can be realized using GPU
technology and these MC methods are labeled as such in column 1 of
Table~1.  The GPU method may be further optimized in combination with
the Lucy (GPU+Lucy) or the B\&W method (GPU+B\&W). On our conventional
computer we use a GPU with 256 threads (8 multiprocessor each with 32
cores) clocked at 1.5 GHz.  When compared to a single thread CPU
application clocked at 3 GHz a speed up factor due to vectorization of
60 is realized.  This is below a theoretical expected speed up factor
of 128 because of additional overheads produced by input/output
routines and memory transfers in the GPU machine.

Total run times of the various methods with different numbers of
threads are given in Table \ref{tab:comp_MC_opt}. The vectorized Lucy
method scales slightly better with the number of available threads
than the other algorithms because fewer atomic memory operations are
required. The speed increase by vectorization as compared to scalar MC
treatments is proportional to the number of available threads. The GPU
method with B\&W optimization is a factor 90 faster than the original
Lucy method.

\section{Ray--tracer to compute SEDs and images}
\label{sec:raytracing}

Photons which eventually escape the MC model space in a particular
solid angle can be counted and converted into a flux density.  In
principle this method allows computing images of the object.
Unfortunately to obtain a moderate signal--to--noise ratio the number
of photon counts per solid angle must be high which is difficult to
reach within reasonable computing times. Here we present a
ray--tracing method which allows computing of noise free images, SEDs
and visibilities at high spatial resolution. The ray--tracer uses the
temperature and scattering events of the cells calculated with the MC
code (Sect.~\ref{sec:MCRT}). The uncertainty in the derived images is
therefore based on the precision of the MC computation.

In the algorithm rays are traced through the model space from an
observer plane of arbitrary orientation.  The ray tracer together with
the MC method allows us to calculate the flux received on each pixel
of an image in the plane of the observer. The observed image is
located at distance $D$ from the object which is at the center of the
cube. The orientation of the image plane is defined by its surface
normal $\vec{~e_{z}}'$. The axis of the image plane
$\vec{~e_{x}}',\vec{~e_{y}}'$ is perpendicular to $\vec{~e_{z}}'$. The
coordinates $(x,y,z)$ of the 3D model space are transformed by a
parallel projection into the coordinates $(x',y')$ of the 2D image in
the observers plane. The image consists of $n_x \times n_y$ pixels,
with a pixel size chosen so that the complete model fits in the
projection.  The center of the image $\vec{i}_0$ is located at
position $[x_0,y_0,z_0]$ and has image coordinates $[x'_0,y'_0]$.  The
projected coordinates $[x',y']$ of the image are transformed into MC
cube coordinates by

\begin{equation}
  \vec{r}(x',y') = \vec{i}_{0}+(x'-x'_{0})\vec{~e_{x}}' + (y'-y'_{0})\vec{~e_{y}}' ~,
\end{equation}
where $\vec{r} = [x,y,z]$. The ray--tracer follows the line of sight
from each detector pixel in direction $\vec{~e_{z}}'$ through the MC
model. Adding the contribution of emission and scattering from each
cell $i$ along the line of sight results in the observed intensity.
The contribution of cell $i$ to the total intensity is

\begin{equation}
  I_{\nu,i} = (I^e_{\nu,i} + I^s_{\nu,i})~e^{-\tau_{\nu,i}} ~,
  \label{eqn:ray1}
\end{equation}
where $I^e_{\nu,i}$ refers to the intensity of the dust emission and
$I^s_{\nu,i}$ to scattering; $\tau_{\nu,i}$ is the optical depth at
frequency $\nu$ of cell $i$. For convenience we drop the index $i$ in
the following. The optical depth is

\begin{equation}
  \tau_{\nu} = \mathrm{K}^{abs}_{\nu} \int_0^{\ell} \rho\,(\vec{r})\, ds ~,
  \label{eqn:ray2}
\end{equation}

where $K^{abs}_{\nu}$ is the absorption cross section per unit dust
mass and $\ell$ is the path lengths. We call the optical depth
measured along the ray from the border of the cell to the observer
$\tau_{\nu}$, and the one measured face--to--face from one side of the
cell to the other $\tau^{\rm cell}_{\nu}$. The emission is
\begin{equation}
  I_{\nu}^{e}=  \big [ 1-\rm{exp}(-\tau^{\rm cell}_{\nu}) \big ]~\mathrm{B_{\nu}}(T) ~.
  \label{eqn:ray3}
\end{equation}


\begin{figure*} [htb]
  \includegraphics[width=\textwidth]{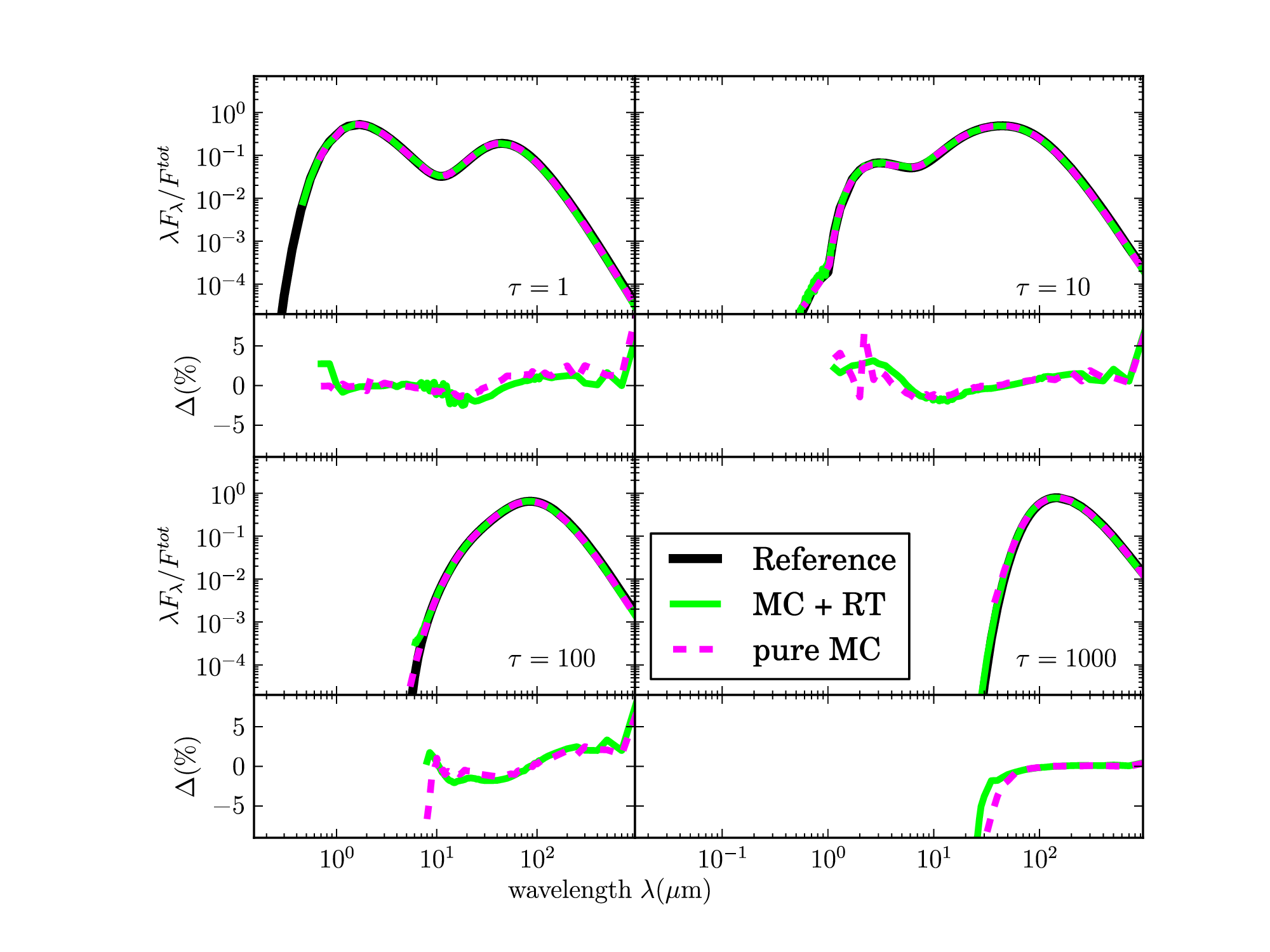}
  \caption{Comparison of SEDs of dust spheres at optical depths
    between $\tau_{\rm{V}}=1$ and 1000 computed with MC and
    ray--tracer (green) or photon counting (dashed), benchmark results
    in black (\cite{ive97}). Lower panels give the difference between
    benchmark and the procedures of this work.  \label{fig:bench1d}}
\end{figure*}

For scattering we implement either a g--factor approximation following
the notation of \cite{kru08} or consider non--isotropic scattering
applying the Henyey--Greenstein phase function $P_{\rm HG} $
\citep{HG41} defined as:

\begin{equation}
  P_{\rm HG}(\beta) = \frac{1}{4\pi}\frac{(1-g^2)}{~\big [ 1+g^2-2g~\rm{cos}(\beta)\big ] ^{3/2}},
\end{equation}
where g is the anisotropy parameter. If $g=0$ the scattering is
isotropic.  The scattered light intensity is given by
\begin{equation}
  I^{s}_{\nu}= \sum_{\beta} P_{\rm HG}(\beta)\frac{N(\beta,\nu)^{sca}~\epsilon }{\Delta\nu} ~,
  \label{eqn:ray4}
\end{equation}


where $\Delta\nu$ is the width of the frequency bin,
$N(\beta,\nu)^{sca}$ is the total number of scattering events with
frequency $\nu$ and angle $\beta$ between the observer and the
original scattering direction. This method to calculate the scattered
intensity is similar to the peel--off technique proposed by
\cite{yus84}. The difference lies in the time of the calculation of
the scattering intensity. The peel--off technique calculates the
intensity during the MC run. Whenever a scattering event occurs the
scattered intensity of the event is added to the observed
images. Therefore in the peel--off technique the location of the
observer has to be known before the MC run. This information is not
required in our MC scheme which stores only the scattering
events. Then, in post--processing using the ray--tracer, the
contribution of the scattered light flux can be computed for arbitrary
observer orientations. The flux density is calculated summing up the
contribution of each cell given by

\begin{equation}
  F_{\nu,\mathrm{i}} = \frac {A_{\rm pix}}{D^{2}} \big [ I_{\nu}^{e}+ I^{s}(\theta) \big ]\ \rm{exp}(-\tau_{\nu}) ~,
\end{equation}
where $\tau_{\nu}$ is the optical depth from the border of the cell to
the observer and $A_{\rm pix}$ is the area of the detector pixel. The
ray--tracer calculates emission and scattering images at high
signal--to--noise. It cannot remove the noise which is introduced by
the MC simulation where the temperature distribution and scattering
events are calculated with a stochastic sampling. However, for a known
distribution of the temperature and scattering events the ray--tracer
provides noise free images.

\begin{figure*} [htb]
  \includegraphics[width=\textwidth]{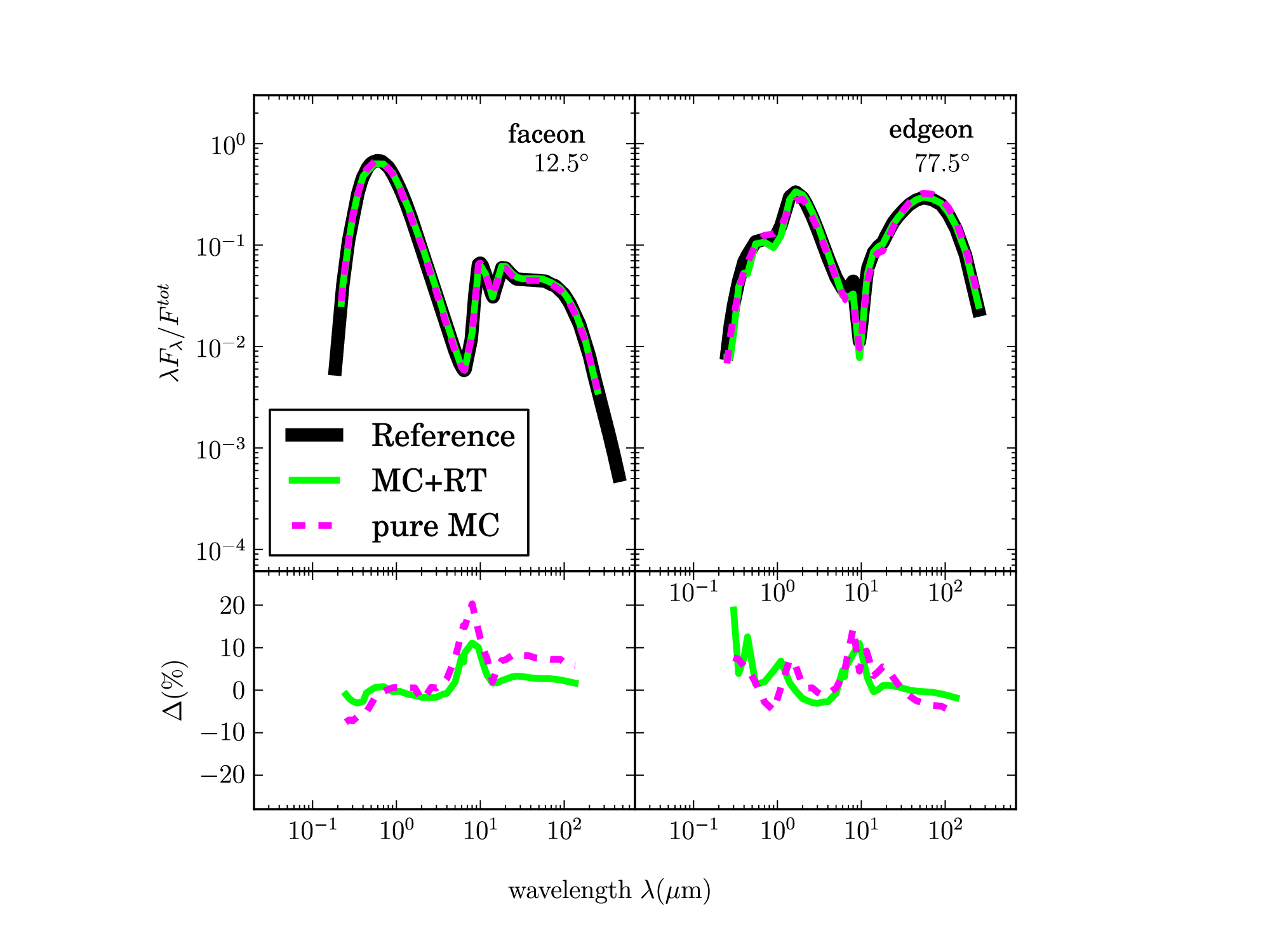}
  \caption{Comparison of SEDs of a dust disk at optical depths along
    the midplane of $\tau_{\rm{V}}=100$, face--on view
    ($12.5^{\circ}$) is shown left and edge--on view ($77.5^{\circ}$)
    right.  SEDs are computed with MC and ray--tracer (green) or
    photon counting (dashed). The mean SED of the benchmark results by
    \cite{pas04} is in black.  Lower panels give the difference
    between the mean SED of the benchmark and the procedures of this
    work. \label{fig:bench2d-1}}
\end{figure*}

\section{Benchmark}
\label{sec:benchmark}

The vectorized MC method is tested against the 1d ray tracing code by
\cite{kru08}. Further verifications of the MC and comparison with
benchmark tests are described below
\subsection{Spherical symmetry (1D)}
\label{sub:bench1d}

Benchmark models in 1D are provided by \cite{ive97}. They consider a
spherical dust envelope which is centrally heated by a 2500\,K star
with luminosity of $1 L_\sun$. For this test the dust absorption and
scattering coefficient for wavelength below $\lambda_0 = 1\mu$m are
$q_{\rm {abs}}=q_{\rm {sca}}=1$ and are $q_{\rm {sca}} =
\lambda_0/\lambda$ and $q_{\rm {abs}} = (\lambda_0/\lambda)^{-4}$ at
longer wavelengths. Models are computed at optical depths of
$\tau_{\rm {V}} = 1$, 10, 100 and 1000 with inner radius $r_{\rm{in}}
= 3.14$, 3.15, 3.20 and 3.52\,AU, respectively. The dust density is
constant and the outer radius is $r_{\rm{out}} = 1000 \times
r_{\rm{in}}$. The models are run on a grid with $10^6$ cells and
$2\times 10^8$ photon packets per iteration are launched from the
star. The number of MC iterations are 3, 4 and 5 for $\tau_{\rm{V}}
\le 10$, 100 and 1000, respectively.

MC computed temperatures are compared to the one calculated by DUSTY
\citep{ive97b} and \cite{kru08}; they agree within $\simless
1$\,\%. The SEDs are shown in Fig. \ref{fig:bench1d} in which each
quadrant represents models at different optical depths.  In the
quadrants the SED are shown in the top and the flux difference between
MC and benchmark in the bottom panel.  The SED of the MC models are
either computed by photon counting or using the ray--tracer
(Sect.~\ref{sec:raytracing}). The difference of the SED computed by MC
and DUSTY is typically better than a few \%. It becomes larger at
faint fluxes where the ray tracing is more accurate than the packet
counting procedure because of photon noise.  For the $\tau_{\rm{V}} =
1000$ model the counting method is starved at fluxes below 0.2\% from
the peak flux. We encountered numerical inconsistencies in the DUSTY
code at flux levels $\simless 10^{-8}$ from the peak when compared
against the radiative transfer code by \cite{kru08}.

The differences between the reference codes and our MC program is
mainly caused by Poisson noise. The photon noise is large at flux
levels which are low compared to the peak flux.  The wavelength range
where the frequency grid of the source and that of the dust overlap is
an additional source of systematic error. The source grid is build so
that the energy of the photon packets of each frequency bin are
identical. This results in a larger bin size at the Rayleigh--Jeans
tail of the star where, on the other hand, the dust grid has a fine
sampling because of the silicate and PAH features.  To reduce this
sampling problem it is necessary to increase the number of frequency
bins of the star. Unfortunately, the total number of photons emitted
from the star must be increased to achieve a certain signal--to--noise
for each bin. The number of photons emitted from the star in each
frequency bin must be large enough to reduce the Poisson noise to
acceptable levels. We use $1000$ frequency bins for the star and emit
in each bin $10^5$ photons. This choice is based on a run time
optimization for the achieved accuracy.  The third source of
systematic differences between the reference codes and our MC scheme
is the spatial grid. While the MC code uses a Cartesian grid in three
dimensions with one level of refinement the reference codes and in
particular that of \cite{kru08}, uses an extremely optimized grid to
account for the spherical symmetry. This systematic effect could be
reduced by an increase of the number of MC grid cells. Current
hardware limits the number of grid cells because of memory and speed
constraints.

\subsection{Disk geometry (2D)}
\label{sub:bench2d}

\begin{table}[ht]
  \caption{Parameters of 2D benchmark by \cite{pas04}}
  \label{tab1}
  \centering
  \begin{tabular}{c l l}
    \hline\hline
    Symbol & Parameter & Value \\
    \hline
    $\rm{M}_{\star}$ & Stellar Mass   & $ 1\ \rm{M}_{\sun}$ \\
    $\rm{R}_{\star}$ & Stellar Radius & $ 1\ \rm{R}_{\sun}$ \\ 
    $\rm{T}_{\star}$ & Stellar effective temperature &$ 5800 \rm{K}$ \\ 
    $r_{\rm {in}}$ & Inner disk radius & $ 1\ \rm{AU}$ \\ 
    $r_{\rm {out}}$ & Outer disk radius & $ 1000\ \rm{AU}$ \\ 
    $r_{d}$ & Half of outer radius & $ 500\ \rm{AU}$ \\
    $z_{d}$ & Disk height & $ 125\ \rm{AU}$ \\
    $\rm{a}$ & Grain radius & $ 0.12\ \mu \rm{m}$ \\ 
    $\rm{\rho}_{0}$ &  Density for $\tau_{\rm{V}}=1$ & $ 8.45\,10^{-22} \rm{g}\rm{cm}^{-3}$ \\ 
    $\rm{\tau}_{\rm{v}}$ & Optical depth at 550\,nm & $ 1, 10, 100$ \\ 
    \hline
  \end{tabular}
\end{table}

Different methods to compute the RT in a 2D dust configuration are
compared by \cite{pas04}.  They consider a star which heats a dust
disk.  The inner region $r < r_{\rm {in}}$ is dust free and the dust
density distribution is similar to those described by
\cite{chi97,chi99}:

\begin{equation}
  \rho(r,z) = \rho_{0}~ (\frac{r_d}{r}) ~ e^{-(0.25\pi(z/h)^{2})} 
\end{equation}
with
\begin{equation}
  h = z_{d} ~ (\frac{r}{r_{d}})^{1.125} ~,
\end{equation}
where $r$ is the distance from the central star in the midplane of the
disc and $z$ is the height above the midplane. The parameter $\rho_0$
is chosen such that the optical depth at 550\,nm along the midplane is
1, 10 and 100, which leads to densities
$\rho_0=8.45\,10^{-22},8.45\,10^{-21}$ and $8.45\,10^{-20}$
g\,cm$^{-3}$.  Additional parameters are specified in
Table~\ref{tab1}. The absorption and scattering cross sections are
that of a 0.12$\mu$m silicate grain (optical data\footnote{download
  from \\http://www.mpia.de/PSF/PSFpages/RT/benchmark.html} are taken
from \cite{dra84}).

The SEDs calculated by the various algorithms and treatments used by
\cite{pas04} agree to better than 20\%. The RT in the disk is solved
by our GPU code and the derived SED are compared to an averaged SED of
the results given by \cite{pas04}

In the MC program we use $~3\times10^6$ cells and run models with
$~2\times 10^8$ photon packets per iteration.

For the optical depth along the midplane of the disk with
$\tau_{\rm{V}} \leq 10$ we use 3 iterations and found in the SED an
overall agreement to the benchmark results to within a few \%. In the
$\tau_{\rm{V}}=100$ case 4 iterations are used and the SED and
differences to a mean SED over those computed in the reference models
is shown in Fig. \ref{fig:bench2d-1}. Two different inclinations
$12.5^\circ$ (face--on), $77.5^\circ$ (edge--on) are displayed. The
residual between SED computed by us and the mean SED of the benchmarks
is typically a few \% and larger deviation are found at fluxes below
1\% of the peak flux.  This is visible at the border of the wavelength
grid and in the 10$\mu$m silicate absorption band where variations
from one SED to another in the reference codes are also larger.

The differences in the two dimensional benchmark and our code are
twofold. There is the Poisson noise which is present in our as well as
in the reference. Note that the reference SED in \cite{pas04} is
computed as an average of various SEDs computed by different
codes. The second source of systematic error is the grid.  We use a
three dimensional Cartesian grid with one level of refinement. The
reference codes calculate the benchmark in a two dimensional grid
optimized to account for the disc symmetry. It is possible to improve
our solution by increasing the number of grid cells as well as the
number of photon packages. However, the current hardware gives an
upper limit to run time and memory requirements.

\begin{figure} [htb]
  \includegraphics[width=0.5\textwidth]{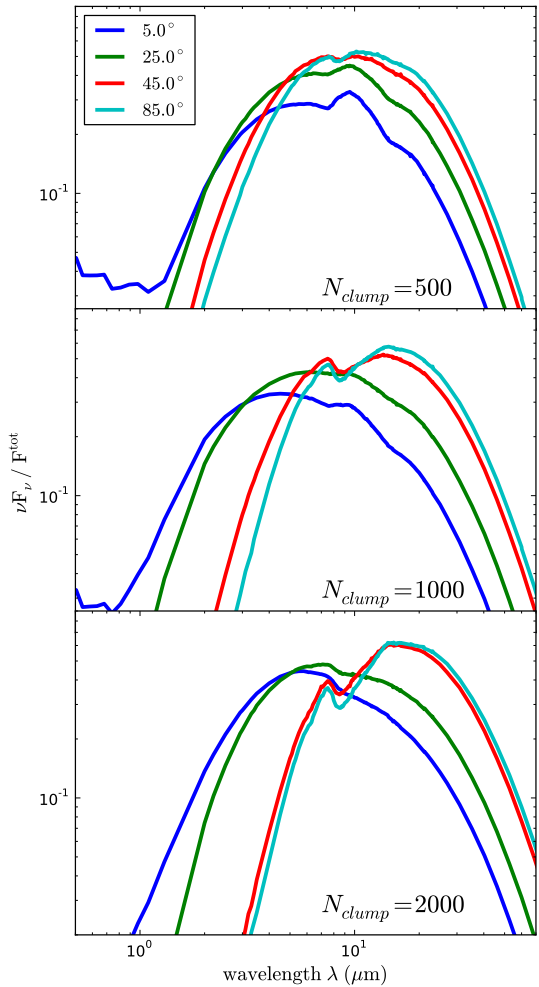}
  \caption{Spectral energy distribution of clumpy AGN torus models at
    viewing angles between $5^\circ$ and $85^\circ$ and number of
    clouds of $N_{\rm{clump}}=500$, 1000 and 2000.}
  \label{fig:sil_clump}
\end{figure}

\begin{figure*}[htb]
  \includegraphics[width=\textwidth]{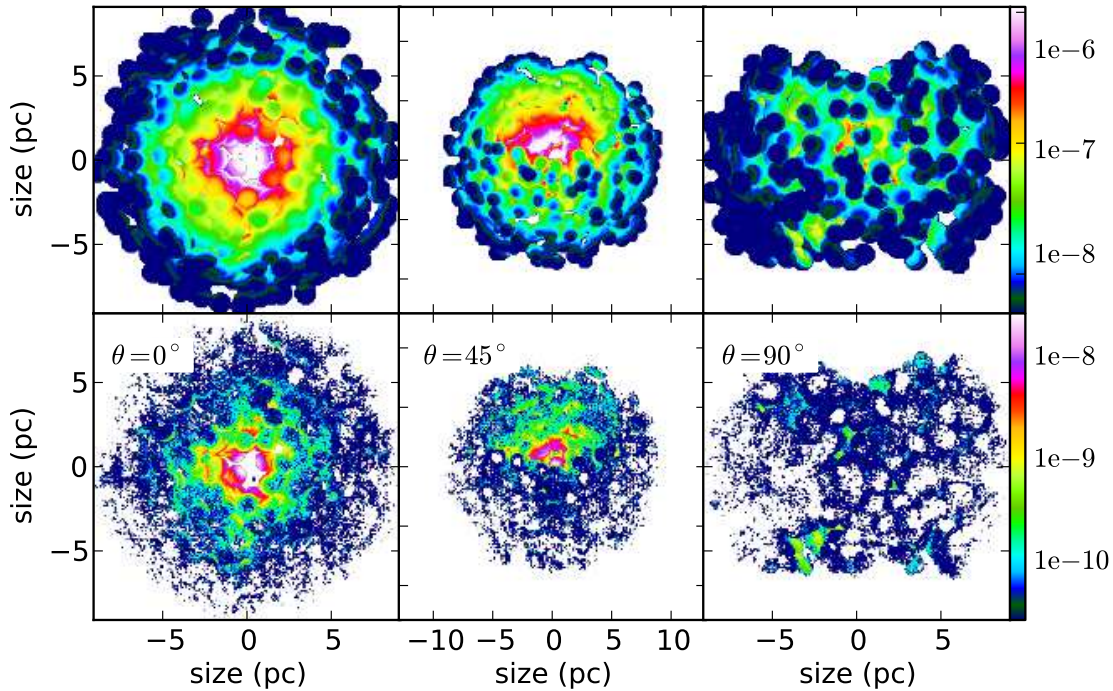}
  \caption{Images at 10$\mu$m (top) and 0.55$\mu$m (bottom) of the
    clumpy AGN torus with $N_{\rm{clump}}=1000$ and viewing angles
    (from left to right) of $0^\circ$, $45^\circ$ and
    $90^\circ$. . Color bar gives flux in (Jy/arcsec$^2$). The source
    with luminosity L=10$^{45}$\,erg/s is at a distance of 15\,Mpc}
  \label{fig:map1000}
\end{figure*}

\section{Clumpy AGN torus model}
\label{sec:clumpy_torus_model}

The unified scheme of AGN \citep{ant85,pad90} predict that the central
engine is surrounded by dust so that the obscuration depends on the
viewing angle.  The detailed geometrical configuration of dust around
the AGN is still a matter of debate \citep{Sta11}. Observations are
not able to resolve the inner part of the objects.  Theoretical
considerations favor a toroidal and clumpy geometry
\citep{nen02,sch05,hoe06,nen08,sch08,sch10}. Observational hints of a
clumpy structure in which optically thick dust clouds surround the
accretion disc of the central black hole are provided by the 10$\mu m$
silicate band. This band is detected in absorption and emission. The
strength of the feature is much weaker than predicted by homogeneous
torus models \citep{pie92,gra94,efs95,vbe03}. A silicate emission
feature is also seen in obscured AGN, where the broad emission lines
are hidden \citep{stu06,mas09}. Observations are explained by
postulating a clumpy torus structure.

For the dust around AGN we consider fluffy agglomerates of silicate
and carbon as sub--particles. We use a power law size distribution:
$n(a) \propto a^{-3.5}$ with particle radii between $a_- \leq a \leq
a_+$ and $a_-=160 \rm{\AA}$\,, $a_+=0.13\mu$m. The bulk density of
both materials is 2.5\,g/cm$^3$. Dust abundances (ppm) are 31 [Si]/[H]
and 200 [C]/[H], which agree with cosmic abundance constraints
\citep{asp09}. This gives a gas--to--dust mass ratio of 130. We take
the relative volume fractions in composite grain to be 34\% silicates,
16\% carbon and 50\% vacuum, which translates, with abundances as
above, to a relative mass fraction of 68\% silicates and 32\%
amorphous carbon. Absorption and scattering cross--sections and the
scattering asymmetry factor is computed with Mie theory and the
Bruggeman mixing rule. Optical constants for silicates are taken from
\cite{dra03} and for carbon we use the ACH2 mixture from
\cite{zub96}. This gives a total mass extinction cross section in the
optical (0.55$\mu$m) of $K^{ext}_V = 35000$.  The wavelength
dependence of $K^{ext}$ is displayed in \cite{kru94}.

The hard (UV/optical) AGN radiation emerges from the accretion disc
around the massive black hole.  Its spectral shape is suggested to
follow a broken power law \citep{row95}:
\begin{equation}
  \lambda F_\lambda \propto \left\{
    \begin{array}{ccl}
      \lambda^{1.2}  & \text{for} & ~-\infty <~ \lambda ~< 0.01\mu {\rm m}\\
      \ {\rm {const}}  & \text{for} & ~~0.01   <~ \lambda ~< 0.10\mu {\rm m}\\
      \lambda^{-0.5} & \text{for} & ~~0.1~~  <~ \lambda ~< ~~1.0\mu {\rm m}\\
      \lambda^{-3}  & \text{for}  & ~~1.0~~ <~ \lambda ~< +\infty\\
    \end{array}
  \right.
\end{equation}

In the following we consider a clumpy AGN dust torus of total
luminosity of $L_{45} = 10^{45}$ erg/s.  The inner radius of the torus
$r_{\rm{in}}$ is set by dust evaporation, which is approximated by
$r_{\rm {in}} = 0.4~(L_{45})^{0.5}\,\rm{pc}$. A clump is represented
as a sphere of constant density and radius of 0.5\,pc.  The optical
depth through the center of the clump is $\tau_{\rm{V}}=30$. The
clumps are randomly distributed within a half opening angle of
$\theta=45^{\circ}$ between an inner radius of $r_{\rm {in}}=0.4$\,pc
and outer radius of $r_{\rm {out}}=6$\,pc.  For random numbers
$\zeta_1,\zeta_2,\zeta_3$ the position of the center of the clouds is
computed by
\begin{eqnarray}
  r &=& \zeta_1~(r_{\rm {out}}-r_{\rm {in}}) \nonumber\\
  \phi &=& 2\pi~\zeta_2\\
  \theta &=& \frac{\pi}{2}+(2~\zeta_3-1)\theta_1\nonumber ~.
\end{eqnarray}
If clumps overlap the density is constant and the same as for a single
clump.

\subsection{Influence on clumps on SED}

The AGN torus model is used to study the influence of clumps on the
SED.  In the models the number of clouds is varied using
$N_{\rm{clump}}=500,1000$ and $2000$. This gives a total optical depth
along the midplane of the torus of $\tau_{\rm{V}} \sim 80$, 140 and
200.  The SED is displayed in Fig.~\ref{fig:sil_clump}; at wavelengths
$\simless\,1\,\mu\rm{m}$, the flux is emitted by the central source
and between 1$\mu$m and 50$\mu$m the SED is dominated by dust
emission. The dust temperatures are between 100 and 1500\,K. At short
wavelengths of the SED the AGN is visible only in the face--on view
and becomes faint at higher inclination angles. At longer wavelengths
the flux is stronger in the edge--on view as compared to observations
at smaller inclination angles. The optical depth of a single clump is
high and already one or two of such intervening clouds define most of
the spectral behavior of the torus. The SEDs at viewing angles between
$45^{\circ}$ and $85^{\circ}$ are similar. At these angles a few
clouds are always in the line of sight and dominate the spectrum
(Fig. \ref{fig:sil_clump}).

Depending on the viewing angle and the total optical depth and dust
temperatures in the AGN, the 10\,$\mu$m silicate band is observed
either in emission or in absorption. In AGN viewed face--on (type I
sources) there is a direct view on the inner region of the torus and a
weak $10 \mu$m silicate emission feature is observed
\citep{sie05,hao05,stu05}. In edge--on sources (type II) the $10 \mu$m
silicate band is seen in absorption.

The silicate feature trace the amount of dust for a given viewing
direction and constraints the distribution of the clumps in the torus
\citep{nik09}. Homogeneous AGN torus models without a clumpy structure
over--predict the strength of the silicate feature \citep{efs95}.  In
the clumpy torus models presented in Fig.~\ref{fig:sil_clump} the $10
\mu$m feature is in emission for face--on views as long as
$N_{\rm{clump}} \simless 1000$.  Individual clouds are optically thick
at $10\mu$m so that a single cloud already obscures the warm and
optical thin region required to observe the silicate band in
emission. This is the case in the face--on view of the model with 2000
clumps where the emission and self--absorption of the silicate grains
cancels out and the spectrum becomes featureless. For inclination
angles $\theta \geq 45^{\circ}$ the silicate features is in absorption
\citep{roc91}. The strength of the absorption feature depends on the
optical depth and therefore increases with increasing number of
clumps.

\subsection{AGN images}
\label{sub:torus-intensity-maps}

The MC combined with the ray--tracer (Sect.~\ref{sec:raytracing})
allows computing of images of the object at a given wavelength.  In
Fig. \ref{fig:map1000} we present a $10\mu$m and $0.55\mu$m image of
the clumpy AGN torus using $N_{\rm{clump}} = 1000$. The mid IR image
is dominated by dust emission and the optical image by scattered
light. The face--on image in the mid IR provides an unobscured view of
the emission from hot dust in the inner torus close to the AGN.  In
this image, clouds near the center dominate the emission. Further out,
clumps become cooler and the contribution to the emission decreases.
In the tilted view, at $45^{\circ}$, intervening clumps obscure part
of the emission from the central region and the contribution of cooler
dust becomes more important than in the face--on view. In the edge--on
view most of the central region is no longer visible and the emission
is dominated by clouds located close to the border of the torus.
However, close to the edge--on view and because of the clumpy nature
of the medium there are still a few lines of sight penetrating to the
inner torus. This is not the case in AGN models using a homogeneous
dust distribution. For all viewing angles and clump configurations the
scattered light images appear similar to the emission images
(Fig.~\ref{fig:map1000}). However, the flux in the optical is two
orders of magnitude fainter than at 10$\mu$m. The scattering light
image in the edge--on view becomes fuzzy. The assumption of isotropic
scattering likely over predicts the scattered light in the face--on
direction. However, the clumpy structure visible in the optical image
is preserved.

\begin{figure}
  \includegraphics[width=0.5\textwidth]{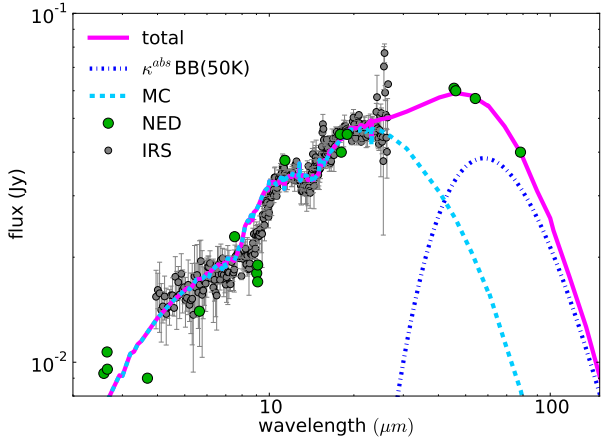}
  \caption{IR emission of the quasar 3C249.1. Photometric data by NED
    and Spitzer spectrum by \cite{sie05}.  The total flux (full line)
    is given as sum of the clumpy AGN torus model (dashed) and a
    cirrus component (dashed dotted).}
  \label{fig:fit}
\end{figure}

\subsection{Quasar 3C 249.1}
\label{sub:3c249.1}

The clumpy AGN dust torus model is applied to fit the SED of the
quasar 3C249.1. The luminosity of the object is $7 \times
10^{45}$\,erg/s at redshift of $z=0.311$ which translates, using
standard cosmological parameters, to a luminosity distance of
1600\,Mpc. The torus inner edge is set by the sublimation temperature
and leads to an inner radius of $r_{\rm {in}}=1$\,pc.  In the inner
region of the model at $r \le 25 \times r_{\rm {in}}$ the half opening
angle of the torus is $\theta_1 = 25^{\circ}$ and in the outer region
up to $r_{\rm {out}} = 50 \times r_{\rm {in}}$ it is $55^{\circ}$.
The torus includes 5500 clouds. The number of clumps per volume
element decreases in the inner part and increases in the outer part of
the torus; other parameters remain the same.

In order to fit data at $\lambda > 40\mu$m we increased the outer
radius of the torus $r_{\rm {out}}$ and considered a two phase medium
in which the density in between the clumps is varied. However, the far
IR emission of the quasar cannot be explained by the AGN torus. The
far IR is dominated by emission of cold dust located furtherout in the
host galaxy. For simplicity we approximate this cirrus component by a
modified black body, $\kappa_{\nu}^{abs}B_{\nu}(T)$ at 50\,K. The dust
torus model fits the silicate emission feature as observed in the
Spitzer spectrum and together with the cirrus component a good fit to
the overall SED is achieved (Fig.~\ref{fig:fit}).

\section{Conclusions}
A parallel 3D MC method is presented to solve the radiative transfer
problem of dust obscured objects in arbitrary geometries.  The code
uses an adaptive three dimensional Cartesian grid. It utilizes the
advantage of different optimization algorithms: in optical thin cells
the Lucy algorithm and in cells at very high optical depths a modified
random walk procedure by \cite{fle84} is included. The temperature of
the dust is computed in an iterative MC method or with the
iteration--free \cite{bw01} algorithm.

The spectral energy distribution of the objects is derived by a photon
counting procedure or a ray--tracing routine. Photons which eventually
escape the MC model space are counted and converted into a flux
density. This method computes the SED of an object very quickly when a
large aperture of several degree is used. Unfortunately, for high
spatial resolution the method is starved by the low count rate. The
SED of an object observed with a pencil beam is computed by a ray
tracer which uses the dust scattering and absorption events of the MC
cells as input. The ray tracer allows computing of noise free images
of the scattered and emitted radiation at any frequency.

MC schemes are known to be computing time expensive. Therefore we
apply particular attention to solve this problem. We take advantage of
the independence of the trajectories of each photon packet.  This
allows a highly vectorized design of the MC program which is not
realized in previous work. With the vectorized code a speed
improvement of about 100 is reached on a low budget computer when
compared to a scalar, non--parallel version of the same program. This
gain in the speed performance of the computations is reached by
applying either a multi--core application using conventional central
processing units (CPU) or the recent technological development of
graphics power units (GPU). The parallelization capabilities of
various MC radiative transfer algorithms are analyzed. By combining
different MC algorithms we report a linear scaling of the computing
time with the number of available threads (processor cores).  The
ray--tracer to compute SED and images is another computer time
expensive procedure and is developed, similar to the MC code, in
vectorized form.

The 3D MC is tested against the ray tracer solution of the radiative
transfer problem in spherical symmetry by \cite{kru08}. Further we
verified our procedure against one and two dimensional benchmarks of
the literature. The code reproduces the spectral energy and dust
temperature distributions of the test cases at high accuracy. As a
first astrophysical application we use the MC code to investigate the
appearance of a dusty AGN at optical and IR wavelengths; a second
application on proto--planetary disks is presented in an accompanying
paper \citep{sie11b}

The dust around the AGN is geometrically configured in a clumpy torus
structure and is represented by fluffy composite grains of various
sizes made of silicate and carbon. The influence of the number of
clumps in the torus on the SED is studied. We find that the clumpy
torus explains the observed ``weak'' silicate absorption and emission
feature observed in AGN.  Images of the AGN in the optical and the mid
IR are presented by viewing the torus from different angles. The
spectral energy distribution of the radio loud quasar 3C249.1 is fit
by the torus model including a cirrus component for the far IR and
submillimeter emission.
\begin{acknowledgements} {We are grateful to Endrik~Kr\"ugel for
    discussions and helpful suggestions. This work is based in part on
    observations made with the {\it Spitzer Space Telescope}, which is
    operated by the Jet Propulsion Laboratory, California Institute of
    Technology under a contract with NASA. This work was partially
    supported by NASA and NSF.}
\end{acknowledgements}

\bibliographystyle{aa} \bibliography{arxiv}

\end{document}